\documentclass[10pt,twocolumn]{IEEEtran}
\usepackage{epsfig}
\usepackage{graphicx}  
\usepackage{cite}
\usepackage{color}
\usepackage{amsmath,amsfonts,amssymb,amscd,amsthm}
\usepackage[dvips]{hyperref}
\usepackage{authblk}
\usepackage{float}
\usepackage{times}
\hypersetup{
	pdfauthor={Yosef Akhtman, R.G. Maunder and Lajos Hanzo},
	pdftitle={An Approximate Coding-Rate Versus Minimum Distance Formula for Binary Codes}}
\newcommand{\figwidth}{\linewidth} 
 
\begin{document}
\setcounter{page}{1}
\title{{An Approximate Coding-Rate Versus Minimum Distance Formula for Binary Codes}}
\author[1]{Yosef Akhtman}
\author[2]{Robert G. Mounder}
\author[2]{Lajos Hanzo}
\affil[1]{\'Ecole Polytechnique F\'ed\'erale de Lausanne, Switzerland}
\affil[2]{School of ECS, University of Southampton, UK}
%
\maketitle

\begin{abstract}
We devise an analytically simple as well as invertible approximate expression, which describes the relation between the minimum distance of a binary code and the corresponding maximum attainable code-rate. For example, for a rate-$(1/4)$, length-$256$ binary code the best known bounds limit the attainable minimum distance to $65 \leq \tilde d(n=256,k=64) \leq 90$, while our solution yields $d(n=256,k=64)=74.4$.
The proposed formula attains the approximation accuracy within the rounding error, and thus satisfies the condition of $\lfloor d\rfloor \leq \tilde d \leq \lceil d\rceil$, for $\approx97\%$ of $(n,k)$ scenarios, where the exact value of the minimum distance $\tilde d$ is known.
The results provided may be utilized for the analysis and design of efficient communication systems.

\end{abstract}

\newcommand\e{e}
\renewcommand\d{{\tt d}}
\newcommand{\mat}[1]{\boldsymbol{\mathit #1}}
\renewcommand\H{\mathcal H}
\newcommand\etal{{\it et. al.}}
\newcommand\efd{d}
\newcommand{\pa}{a(n)}
\newcommand{\pb}{b(n)}
\newcommand{\pc}{c(n)}
\newcommand{\px}{\xi(n)}
\section{Introduction}
\label{sec:intro}
One of the fundamental open problems in coding theory is constituted by the issue of determining the highest cardinality $|\mathcal C|=2^{k}$ attainable by a binary code $\mathcal C$ of length $n$, having a rate of $r=k/n$ and a minimum distance of $\efd$ \cite{Moon2006}, where the minimum distance $\efd$ is defined as the minimum Hamming distance between any two codewords in the codebook $\mathcal C$. In addition to its theoretical significance, the problem considered appears in numerous important applications, including the design of efficient coding schemes and their characterization in terms of the achievable probability of error. Although the complete solution of the rate-versus-minimum-distance problem does not exist at the time of writing, several theoretical lower and upper bounds on the desired relation may be found in the literature \cite{Hamming1950,Gilbert1952,Plotkin1960,McEliece1977,Moon2006}.
In particular, the tightest known bounding characteristics, which originate from a variety of theoretically, as well as empirically obtained sources \cite{Brouwer98}, are provided by the code-tables compiled by Grassl {\em et. al.} in \cite{Grassl:codetables}.

Specifically, some of the best known asymptotic ($n\rightarrow\infty$) as well as finite-$n$-related lower and upper bounds are summarized in Table~\ref{tab:rate-weight-bounds}, where we define the binary entropy function $H(q)=-q\log_2(q)-(1-q)\log_2(1-q)$ and denote a normalized minimum distance as $\delta=d/n$. 
More specifically, the tightest known asymptotic ($n\rightarrow\infty$) lower bound was derived by Gilbert \cite{Gilbert1952}, while the corresponding upper bounds were devised by Hamming \cite{Hamming1950} and McEliece \etal~(MRRW) \cite{McEliece1977}. 
The prominent asymptotic lower and upper bounds are depicted in Figure~\ref{fig:rate-dmin-asympt}. 
Furthermore, the best known finite-$n$ bounds are constituted by the Gilbert lower bound, as well as the Hamming and Plotkin upper bounds \cite{Plotkin1960}.
The finite-$n$ lower and upper bounds for the specific case of having $n=7$ are depicted in Figure~\ref{fig:rate-dmin-n}.

Unfortunately, however, most of the available theoretical, as well as empirical bounds are notoriously difficult to use in practice. On the one hand, as may be inferred from Figures~\ref{fig:rate-dmin-asympt} and \ref{fig:rate-dmin-n}, the asymptotic bounds provide little information about the desired characteristics of a wide range of finite-$n$ scenarios, routinely encountered in practical applications. On the other hand, the theoretical bounds corresponding to the finite-$n$ cases involve excessively complex numerical computations. Against this background, {\em the novel contribution of this paper is constituted by the formulation of an analytically simple as well as invertible expression $r(n,\delta)$, which complies with all known theoretical bounds in both finite-$n$ and asymptotic ($n\!\!\rightarrow\!\!\infty$) contexts, while accurately approximating the empirical bounds, and thus providing a practical tool for the analysis and design of efficient binary codes.}
{We would like to explicitly emphasise the applied nature of this study, which is aimed at the development of a methodology for the analysis and optimization of communication networks discussed, for example, in \cite{Akhtman2009e}.}

\hspace{-5mm}
\begin{table*}[htb]\footnotesize
  \centering
  \caption{Known bounds on the maximum code rate achievable for a given $(n,d)$ (finite length case) or $\delta$ (asymptotic case).}
  \begin{tabular}{p{20mm}|l l p{40mm}}
    \ & finite $n$ & asymptotic $n\rightarrow\infty$ & notes \\ \hline
    Varshamov-Gilbert \cite{Gilbert1952} & $r \geq \displaystyle 1-\dfrac1n\log_2 \sum_{i=0}^{d-1} {n \choose i}$ & $r \geq 1-H(\delta)$ & tightest known lower bound \\
    Hamming \cite{Moon2006}   & $r \leq \displaystyle 1-\dfrac1n\log_2 \sum_{i=0}^{\lfloor (d-1)/2 \rfloor} {n \choose i}$ & $r \leq 1-H(\delta/2)$ & tight upper bound for very high rate codes\\
    MRRW \cite{McEliece1977} & \ & $r \leq H(1/2-\sqrt{\delta(1-\delta)})$ & tightest known asymptotic upper bound for medium and low-rate codes\\
    Plotkin \cite{Plotkin1960} & $r \leq \dfrac 1n\left[1-\log_2(2-\dfrac 1\delta)\right]$ & \ & very tight upper bound for \mbox{$\delta>1/2$}
  \end{tabular}
  \label{tab:rate-weight-bounds}
\end{table*}

\section{Rate versus minimum distance trade-off}
\label{sec:semi-method-model-bound}
Firstly, let us consider three special cases, where the exact value of the maximum minimum distance $\efd$ is known.
\renewcommand{\labelenumi}{\alph{enumi})}
\begin{enumerate}
  \item For a unity-rate binary code of length $n=1,2,\dots$, we have $\efd=1$.
  \item The {\em simplex code} for block length of $n=2^k-1,\ k=1,2,\dots$ exhibits a rate of $r\!\!=\!\!k/(2^k-1)$ and a constant Hamming distance of $\efd=2^{k-1}$ between any pair of codewords.
  \item For any block length $n=1,2,\dots$, we may consider an optimum rate-$(r\!\!=\!\!1/n)$ $n$-{\em repetition code} conveying a single bit of information and exhibiting $\efd=n$. 
\end{enumerate}

\begin{figure}[htb]
  \begin{center}
    \includegraphics[width=\figwidth]{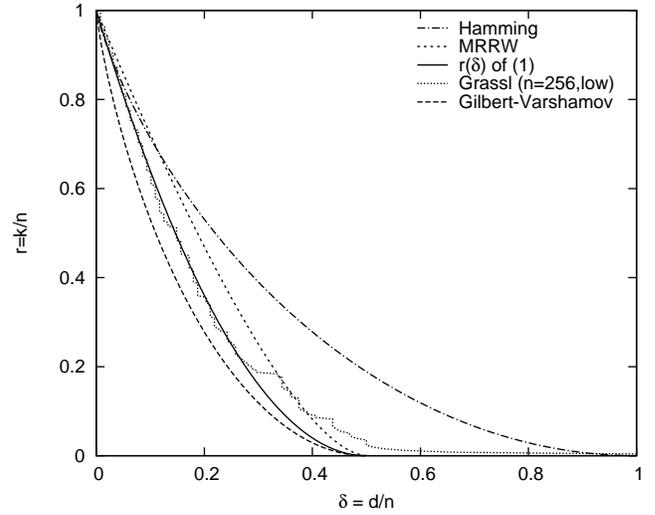}
  \end{center}
  \caption{Rate versus normalized minimum distance for known asymptotic bounds.}
  \label{fig:rate-dmin-asympt}
\end{figure}

\begin{figure}[htb]
  \begin{center}
    \includegraphics[width=\figwidth]{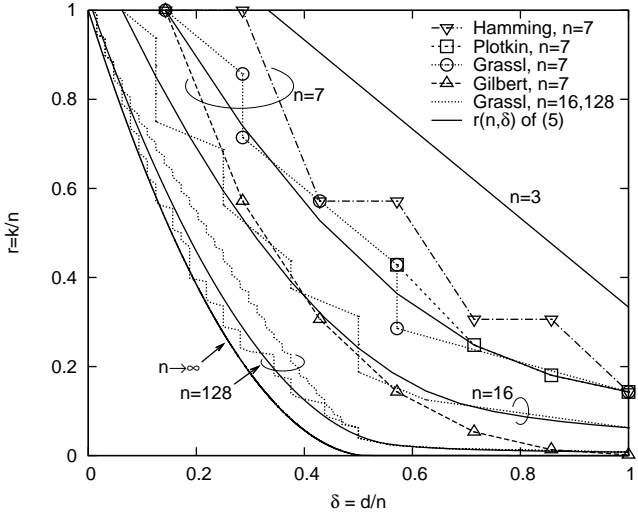}
  \end{center}
  \caption{Rate versus normalized minimum distance for finite length codes.}
  \label{fig:rate-dmin-n}
\end{figure}

Secondly, we would like to point out the following list of important empirical observations.
\renewcommand{\labelenumi}{\roman{enumi})}
\begin{enumerate}
\item As confirmed by Figure~\ref{fig:rate-dmin-asympt}, a simple quadratic function 
  \begin{align}
    r(\delta) = (2\delta-1)^2
    \label{eq:semi-method-model-parab}
  \end{align}
provides an accurate approximation of the empirical lower bound \cite{Grassl:codetables} for the code length of $n=256$ and rates in excess of $0.2$. Notably, (\ref{eq:semi-method-model-parab}) satisfies all known asymptotic bounds, namely the upper MRRW \cite{McEliece1977} and Hamming \cite{Hamming1950} bounds, as well as the lower Gilbert-Varshamov \cite{Gilbert1952} bounds summarized in Table~\ref{tab:rate-weight-bounds}, over the entire range of practically significant code rates\footnote{It should be noted that the expression in~(\ref{eq:semi-method-model-parab}) does not satisfy the Hamming asymptotic upper bound for a hypothetical range of long, very high rate codes $(n>200,r>0.9)$, which exhibit no practical significance due to their low coding gain and excessive decoding complexity.}.
\item As exemplified by the specific case of $n=7$, portrayed in Figure~\ref{fig:rate-dmin-n}, the actual achievable values $r(\delta)$ constitute a discrete function, which cannot have an exact monotonic analytical description.   
\item As may be inferred from comparing Figures~\ref{fig:rate-dmin-asympt} and~\ref{fig:rate-dmin-n}, the asymptotic bounds of Figure~\ref{fig:rate-dmin-asympt} provide little useful information about the desired characteristics of short codes having $1 \leq n \ll 100$, and representing a considerable practical importance in the design of, for example, interactive, real-time speech and video systems.
\item As further suggested by the specific example of having $n=7$, both the finite-$n$ Gilbert and Hamming bounds are relatively loose, while the Plotkin bound is tight for $\delta>{\lceil n/2\rceil}/{n}$.
\item The Plotkin upper bound coincides with the actual achievable maximum rate $r$ in the special cases of $(b)$ and $(c)$ considered above, which further substantiates the assumption that the Plotkin bound constitutes the tightest possible analytical bound in the $\delta>{\lceil n/2\rceil}/{n}$ range.
\end{enumerate}

Taking into consideration observations (i)-(v), we hypothesize a solution exhibiting the following properties:
\renewcommand{\labelenumi}{\arabic{enumi})}
\begin{itemize}
\item Asymptotic quadratic approximation of (\ref{eq:semi-method-model-parab})
  \begin{align}
    \lim_{n\rightarrow\infty}r(n,\delta) = (2\delta-1)^2.
    \label{eq:semi-method-model-const1}
  \end{align}
\item Unity-rate special case (a)
  \begin{align}
    r(n,1/n) = 1.
    \label{eq:semi-method-model-const2}
  \end{align}
\item Plotkin bound \cite{Plotkin1960} and special cases (b) and (c)
  \begin{align}
    r\left(n,\delta>\frac{\lceil n/2\rceil}{n}\right) \approx \frac 1n [1-\log_2(2-1/\delta)].
    \label{eq:semi-method-model-const3}
  \end{align}
\end{itemize}
Specifically, we propose a solution in the form of a smooth two-segment function $r(n,\delta)$ expressed as
\begin{align}
  r(n,\delta) = \left\{ \begin{array}{ l l }
     \!\!a(n)\delta^2+b(n)\delta+c(n)    & \text{if } \delta < {\lceil{n/2+\xi(n)}\rceil}/{n} \\
     \!\!\dfrac 1n[1-\log_2(2-1/\delta)] & \text{otherwise},
  \end{array} \right .
\label{eq:semi-method-model-r-vs-d}
\end{align}
where the free parameters $a,b,c$ and $\xi$ depend on the code-length $n$ and are chosen to ensure that the quadratic constituent in Equation~(\ref{eq:semi-method-model-r-vs-d}) complies with the constraints (\ref{eq:semi-method-model-const1}) and (\ref{eq:semi-method-model-const2}), while the constraint (\ref{eq:semi-method-model-const3}) is automatically obeyed by the corresponding logarithmic constituent of (\ref{eq:semi-method-model-r-vs-d}).

\renewcommand\d\delta
Furthermore, the requirement of smoothness in the expression of (\ref{eq:semi-method-model-r-vs-d}) imposes the following additional constraints on the quadratic constituent in (\ref{eq:semi-method-model-r-vs-d}):
\begin{itemize}
\setcounter{enumi}{3}
\item Continuity at the transition point of\footnote{Here and in the following we use $a,b,c$ and $\xi$ instead of $a(n),b(n),c(n)$ and $\xi(n)$ for the sake of brevity.} $\delta_2\!=\!{\lceil{n/2+\xi}\rceil}/{n}$
  \begin{align}
    r_2 = a\d_2^2 + b\d_2 + c = [1-\log_2(2-1/\d_2)]/n.
    \label{eq:semi-method-model-const4}
  \end{align}
\item Continuity of the first derivative at the transition point $\d_2$, which may be attained by imposing continuity of the discrete function of (\ref{eq:semi-method-model-r-vs-d}) in the next consecutive point $\d_3=(\lceil{n/2+\xi}\rceil+1)/{n}$, yielding
  \begin{align}
    r_3 = a\d_3^2 + b\d_3 + c = [1-\log_2(2-1/\d_3)]/n.
    \label{eq:semi-method-model-const5}
  \end{align}
\end{itemize}
By combining the constraints of (\ref{eq:semi-method-model-const4}) and (\ref{eq:semi-method-model-const5}) with (\ref{eq:semi-method-model-const2}), we arrive at a system of three equations, which uniquely determines the values of the parameters $a,b$ and $c$. Specifically, we have
\begin{align}
  \left\{ \begin{array}{ l }
     r_1 = a\d_1^2 + b\d_1 + c \\
     r_2 = a\d_2^2 + b\d_2 + c \\
     r_3 = a\d_3^2 + b\d_3 + c,
  \end{array} \right .
\label{eq:semi-method-model-eqsys}
\end{align}
where in addition to the parameters defined in (\ref{eq:semi-method-model-const4}) and~(\ref{eq:semi-method-model-const5}), we have $r_1=1$ and $\d_1=1/n$.
The general solution of the system of equations in~(\ref{eq:semi-method-model-eqsys}) is given by
\begin{align}
  a &= \frac{r_3 \left(\d_2-\d_1\right)+r_2 \left(\d_1-\d_3\right)+r_1
    \left(\d_3-\d_2\right)}{\left(r_1-r_2\right) \left(r_1-r_3\right) \left(r_2-r_3\right)},\notag \\
  b &= \frac{\left(\d_2-\d_3\right) r_1^2+r_3^2 \left(\d_1-\d_2\right)+r_2^2
    \left(\d_3-\d_1\right)}{\left(r_1-r_2\right) \left(r_1-r_3\right) \left(r_2-r_3\right)},\notag \\
  c &= \frac{\left(r_3 \d_1-x_1 \d_3\right) r_2^2+\left(r_1^2 \d_3-r_3^2 \d_1\right) r_2+r_1 r_3
   \left(r_3-r_1\right) \d_2}{\left(r_1-r_2\right) \left(r_1-r_3\right)
   \left(r_2-r_3\right)}.
 \label{eq:semi-method-model-abc}
\end{align}
Observe that despite it seemingly complex appearance, Equation~(\ref{eq:semi-method-model-abc}) contains simple closed-form expressions, which may be readily calculated for any given value of $n$. Furthermore, it may be readily demonstrated that constraint (\ref{eq:semi-method-model-const1}) is satisfied if
\begin{align}
  \lim_{n\rightarrow\infty} \xi = \infty
  \label{eq:semi-method-model-xi1}
\end{align}
and
\begin{align}
  \lim_{n\rightarrow\infty} \frac{n/2+\xi}{n} = \frac 12\ \Rightarrow \ 
  \lim_{n\rightarrow\infty} \frac{\xi}{n} = 0.
  \label{eq:semi-method-model-xi2}
\end{align}
Our analysis has shown that any sensible choice of the function $\xi(n)$ as monotonically increasing and satisfying the conditions~(\ref{eq:semi-method-model-xi1}) and~(\ref{eq:semi-method-model-xi2}) as well as $0\leq\xi(1)\leq1$ yields similar results. Specifically, in this study we assume having
\begin{align}
  \xi(n)=\log_2(n)/2.
  \label{eq:semi-method-model-xi3}
\end{align}
Some examples of values of the parameters $a,b,c$ and $\xi$ calculated using Equations~(\ref{eq:semi-method-model-abc}) and~(\ref{eq:semi-method-model-xi3}) for various code-lengths $n$ are summarized in Table~\ref{tab:semi-approx-param}.
%
%

\begin{table}[htb]
  \centering
  \caption{Approximation parameters $a,b,c$ and $\xi$ calculated using~(\ref{eq:semi-method-model-abc}) and~(\ref{eq:semi-method-model-xi3}) for some values of the code-length $n$.}
  \begin{tabular}{c c c c c}
    $n$ & $a$ & $b$ & $c$ & $\xi$    \\ \hline
    $4$ & $0.83$ & $-2.04$ & $1.46$ & $1.0$ \\
    $8$ & $1.23$ & $-2.36$ & $1.28$ & $1.5$ \\
    $16$ & $1.72$ & $-2.70$ & $1.16$ & $2.0$ \\
    $32$ & $2.17$ & $-2.99$ & $1.09$ & $2.5$ \\
    $64$ & $2.57$ & $-3.22$ & $1.05$ & $3.0$ \\
    $128$ & $2.92$ & $-3.42$ & $1.03$ & $3.5$ \\
    $256$ & $3.11$ & $-3.53$ & $1.01$ & $4.0$
  \end{tabular}
  \label{tab:semi-approx-param}
\end{table}

\begin{figure}[htb]
  \begin{center}
    \includegraphics[width=\figwidth]{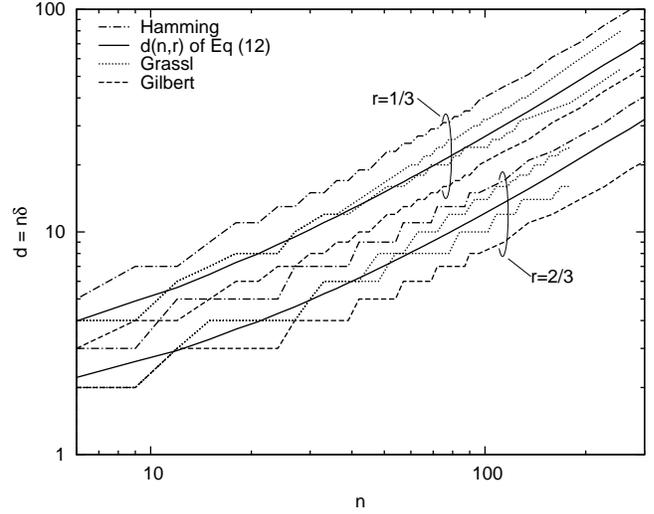}
  \end{center}
  \caption{Minimum distance versus code-length for binary codes of rates $r=1/3$ and $2/3$, as calculated using the theoretical Gilbert and Hamming bounds \cite{Gilbert1952,Hamming1950}, the empirical Grassl bounds \cite{Grassl:codetables}, as well as the proposed expression of~(\ref{eq:semi-method-model-d-vs-k}).}
  \label{fig:rate-dmin-dvsn}
\end{figure}

\begin{figure}[htb]
  \begin{center}
    \includegraphics[width=\figwidth]{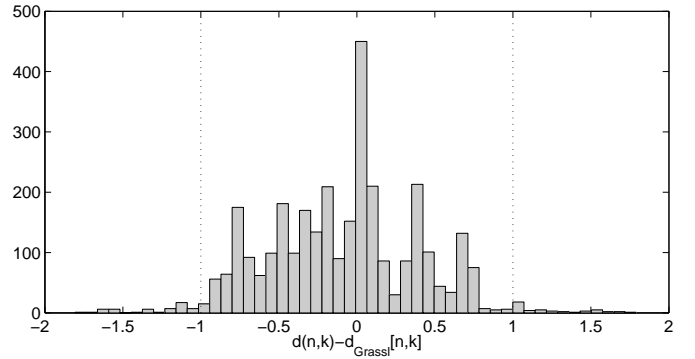}
  \end{center}
  \caption{Histogram of the approximation error $e[n,k]=d(n,k)-d_{\rm Grassl}[n,k]$ based on the 3856 scenarios $(16\leq n\leq 256,\{1\leq k\leq 8,(n-7)\leq k\leq n\})$ for which the exact maximum minimum distance $d_{\rm Grassl}[n,k]$ is known \cite{Grassl:codetables}. The approximation accuracy of $|e[n,k]| < 1$ was achieved in $\approx\!\! 97\%$ of the cases considered, while the accuracy of $1\leq |e[n,k]| < 2$ was achieved in the remaining $3\%$ of the cases.}
  \label{fig:dmin-fithist}
\end{figure}

The resultant expression $r(n,\delta)$ of Equation~(\ref{eq:semi-method-model-r-vs-d}) is compared to the available theoretical and empirical bounds in Figures~\ref{fig:rate-dmin-asympt} and \ref{fig:rate-dmin-n} for the asymptotic case $(n\rightarrow\infty)$, where we have the original quadratic expression $r(n,\delta)\rightarrow(2\delta-1)^2$, and the finite-$n$ cases of $n=3,7,16,128$, respectively. 

Expression (\ref{eq:semi-method-model-r-vs-d}) may be deemed analytically simple, since it has a closed form and is composed of elementary functions. Moreover, (\ref{eq:semi-method-model-r-vs-d}) is readily invertible, yielding
\begin{align}
  \delta(n,r) = \left\{ \begin{array}{ l l }
     \dfrac {-b-\sqrt{b^2-4a(c-r)}}{2a} & \text{if } r > \frac 1n \log_2(n+1) \\
     \dfrac {2^{rn-1}}{2^{rn}-1}        & \text{otherwise},
  \end{array} \right .
\label{eq:semi-method-model-d-vs-k}
\end{align}
where the coefficients $a,b$ and $c$ may be readily calculated using (\ref{eq:semi-method-model-const4})--(\ref{eq:semi-method-model-abc}). In the asymptotic case of having $n\rightarrow\infty$, which in practice may be safely employed for all scenarios having $n\gg100$, we may simply use the inverse of (\ref{eq:semi-method-model-parab}), yielding $\delta(r) = (1+\sqrt{r})/2$.

Figure~\ref{fig:rate-dmin-dvsn} portrays the comparison between the formula of Equation~(\ref{eq:semi-method-model-d-vs-k}) and the best available theoretical and empirical upper and lower bounds for the specific cases of rate-$(1/3)$ and rate-$(2/3)$ binary codes. 
Observe, that the Hamming and Gilbert theoretical bounds imply a considerable ambiguity in terms of the attainable minimum distance $d(n,r)$. Furthermore, the devised expression of~(\ref{eq:semi-method-model-d-vs-k}) provides an accurate approximation of the available empirical Grassl bounds \cite{Grassl:codetables} for both $r=1/3$ and $2/3$ cases. 

The approximation accuracy of the proposed formula of (\ref{eq:semi-method-model-d-vs-k}) was further tested using the 3856 scenarios $(16\leq n\leq 256,\{1\leq k\leq 8,(n-7)\leq k\leq n\})$ for which the exact maximum minimum distance $d_{\rm Grassl}[n,k]$ is known from \cite{Grassl:codetables}. The histogram of the resultant approximation error $e[n,k]=n\delta(n,k)-d_{\rm Grassl}[n,k]$ calculated using (\ref{eq:semi-method-model-d-vs-k}) is depicted in Figure~\ref{fig:dmin-fithist}. Specifically, in approximately $97\%$ of the cases considered, the accuracy of $|e[n,k]| < 1$ was achieved, thus suggesting that the desired value $d_{\rm Grassl}[n,k]$ was the nearest integer higher or lower than the real number $d(n,k)$ provided by the approximation formula. Furthermore, the approximation accuracy of $1\leq |e(n,k)| < 2$ was achieved in the remaining $3\%$ of the cases.

\section{Conclusion}
We formulated an analytically simple as well as invertible expression $r(n,\delta)$, which approximates the optimum trade-off between the maximum rate and the corresponding maximum minimum distance attainable by binary codes of length $n$. The resultant closed-form analytical expression accurately approximates the best available empirical bounds and complies with all known theoretical bounds in both finite-$n$ as well as in asymptotic ($n\rightarrow\infty$) contexts.

For example, for a rate-$(1/4)$, length-$256$ binary code the best known bounds limit the attainable minimum distance to $65 \leq d(n=256,k=64) \leq 90$, while our solution yields $d(n=256,k=64)=74.4$.
The proposed formula attains the approximation accuracy within the rounding error, and thus satisfies the condition of $\lfloor d(n,k)\rfloor \leq d_{\rm Grossl}[n,k] \leq \lceil d(n,k)\rceil$, for $\approx97\%$ of $(n,k)$ scenarios, where the exact value of the maximum minimum distance $\tilde d_{\rm Grossl}[n,k]$ is known. Furthermore, the condition of $\lfloor d(n,k)-1\rfloor \leq d_{\rm Grossl}[n,k] \leq \lceil d(n,k)+1\rceil$ is satisfied in $100\%$ of the cases considered. Correspondingly, the proposed method provides a practical tool for the design and characterization of efficient communication systems.
%

%
\bibliographystyle{IEEEtran} 
\bibliography{comms,publist}
\end{document}